\documentclass[]{spie}  
\usepackage[]{amsmath}
\usepackage[]{amssymb}
\usepackage[]{graphicx}
\usepackage{xspace} 

\def\eg{{\it e.g.}\xspace}
\def\etal{{\it et al.}\xspace}
\def\spider{{\sc Spider}\xspace}

\title{Design and performance of the Spider instrument} 

\author{M.~C.~Runyan\supit{a}, P.A.R.~Ade\supit{b},
M.~Amiri\supit{c}, S.~Benton\supit{d},
R.~Bihary\supit{e}, J.J.~Bock\supit{a,f},
J.R.~Bond\supit{g}, J.A.~Bonetti\supit{f},
S.A.~Bryan\supit{e}, H.C.~Chiang\supit{h}, 
C.R.~Contaldi\supit{i}, B.P.~Crill\supit{a,f}, 
O.~Dore\supit{a,f},
D.~O'Dea\supit{i}, M.~Farhang\supit{d}, 
J.P.~Filippini\supit{a},
L.~Fissel\supit{d}, N.~Gandilo\supit{d},
S.R.~Golwala\supit{a}, J.E.~Gudmundsson\supit{h},
M.~Hasselfield\supit{c}, M.~Halpern\supit{c},
G.~Hilton\supit{j}, W.~Holmes\supit{f},
V.V.~Hristov\supit{a}, K.D.~Irwin\supit{j},
W.C.~Jones\supit{h}, C.L.~Kuo\supit{k},
C.J.~MacTavish\supit{l},
P.V.~Mason\supit{a}, T.A.~Morford\supit{a},
T.E.~Montroy\supit{e}, C.B.~Netterfield\supit{d},
A.S.~Rahlin\supit{h}, C.D.~Reintsema\supit{j},
J.E.~Ruhl\supit{e}, M.C.~Runyan\supit{a},
M.A.~Schenker\supit{a}, J.~Shariff\supit{d},
J.D.~Soler\supit{d}, A.~Trangsrud\supit{a},
R.S.~Tucker\supit{a}, C.~Tucker\supit{b}, and
A.~Turner\supit{f}
\skiplinehalf
\supit{a}Division of Physics, Mathematics, and Astronomy, California Institute of Technology,
Pasadena, CA, USA; \\
\supit{b}School of Physics and Astronomy, Cardiff University, Cardiff, UK; \\
\supit{c}Department of Physics and Astronomy, University of British
Columbia, Vancouver, BC, Canada; \\
\supit{d}Department of Physics, University of Toronto, Toronto, ON,
Canada; \\
\supit{e}Department of Physics, Case Western Reserve University,
Cleveland, OH, USA; \\
\supit{f}Jet Propulsion Laboratory, Pasadena, CA, USA; \\
\supit{g}Canadian Institute for Theoretical Astrophysics, University
of Toronto, Toronto, ON, Canada; \\
\supit{h}Department of Physics, Princeton University, Princeton, NJ, USA; \\
\supit{i}Department of Physics, Imperial College, University of
London, London, UK; \\
\supit{j}National Institute of Standards and Technology, Boulder, CO, USA; \\
\supit{k}Department of Physics, Stanford University, Stanford, CA, USA; \\
\supit{l}Kavli Institute for Cosmology, University of Cambridge, Cambridge, UK }

\begin{document} 
  \maketitle 

\begin{abstract}
Here we describe the design and performance of the \spider instrument.  \spider is a balloon-borne cosmic microwave background polarization imager that will map part of the sky at 90, 145, and 280~GHz with sub-degree resolution and high sensitivity.  This paper discusses the general design principles of the instrument inserts, mechanical structures, optics, focal plane architecture, thermal architecture, and magnetic shielding of the TES sensors and SQUID multiplexer.  We also describe the optical, noise, and magnetic shielding performance of the 145~GHz prototype instrument insert.
\end{abstract}


\keywords{\spider, cosmic microwave background, polarization, cryogenics, magnetic shielding}

\section{INTRODUCTION}
\label{sec:intro}  

\subsection{The \spider Project} 

The \spider instrument in a balloon-borne millimeter-wave polarimeter designed to make very high sensitivity measurements of the polarization in the cosmic microwave background on mid- to large-angular scales with the goal of detecting primordial gravity waves from the inflationary epoch of the early universe.  \spider will observe at 90, 145, and 280~GHz with beam sizes of $51'$, $31'$, and $17'$, respectively.  An Antarctic flight of 24 days will allow \spider to map 8\% of the sky to a depth of 0.25, 0.21, and 0.74~$\mu K_{CMB}$ per square degree at 90, 145, and 280~GHz, respectively. A more detailed description of the \spider project can be found in Filippini \etal \cite{filippini:spie2010}, in these proceedings, as well as Crill \etal \cite{crill:spie2008}.

\subsection{The \spider Instrument Payload and Cryostat} 

At the heart of the \spider balloon payload is a large liquid helium cryostat.  The cryostat houses up to six instrument inserts (described below) bolted to a $\sim1000\ell$ helium tank with all of the inserts pointed in the same direction.  The cryostat cryogenics are briefly described in section \ref{sec:cryogenics} below and a more thorough discussion can be found in Gudmundsson \etal \cite{gudmundsson:spie2010} in this volume.  The cryostat measures 2.2~m in length and 2.0~m in diameter.  The cryostat is mounted to a carbon fiber gondola via two pillow blocks that allow the cryostat to tilt in elevation.  The gondola will be suspended from a 8 million cubic foot helium stratospheric balloon and is steerable in both elevation and azimuth.  Power to the instrument is provided by solar arrays mounted to the back side of the payload and the instrument always points no closer than $45\deg$ to the sun.

\section{Instrument Inserts}

\subsection{Introduction and Design Considerations}

One of the biggest concerns for \spider is instrument payload mass.  The science payload lifting capacity of the balloon is limited to $\sim5000$ pounds and the duration of our flight will depend, in part, upon payload mass.  To that end we have tried to balance the constraint on mass with the desire to build a stiff structure that will not deflect under the periodic acceleration of the gondola scan pattern, nor break during launch (and hopefully landing).  Materials with high thermal conductivity (such as copper) and good magnetic shielding properties (niobium, lead, and high-permeability materials) tend to be dense.  So we have tried to minimize their use where ever possible.  We had the benefit of starting the \spider insert design after much of the design of our sister experiment BICEP2 was completed.  BICEP2 was designed for operation from the ground where weight is not as serious of a concern.  We identified many areas where we could reduce the weight of the insert design with either a change in materials, strategic light weighting, or completely modifying a component (such as changing large metal tubes for carbon fiber trusses sheathed in light-weight copper-clad fiberglass).

\begin{figure}[!t]
\begin{center}
\includegraphics[height=3in]{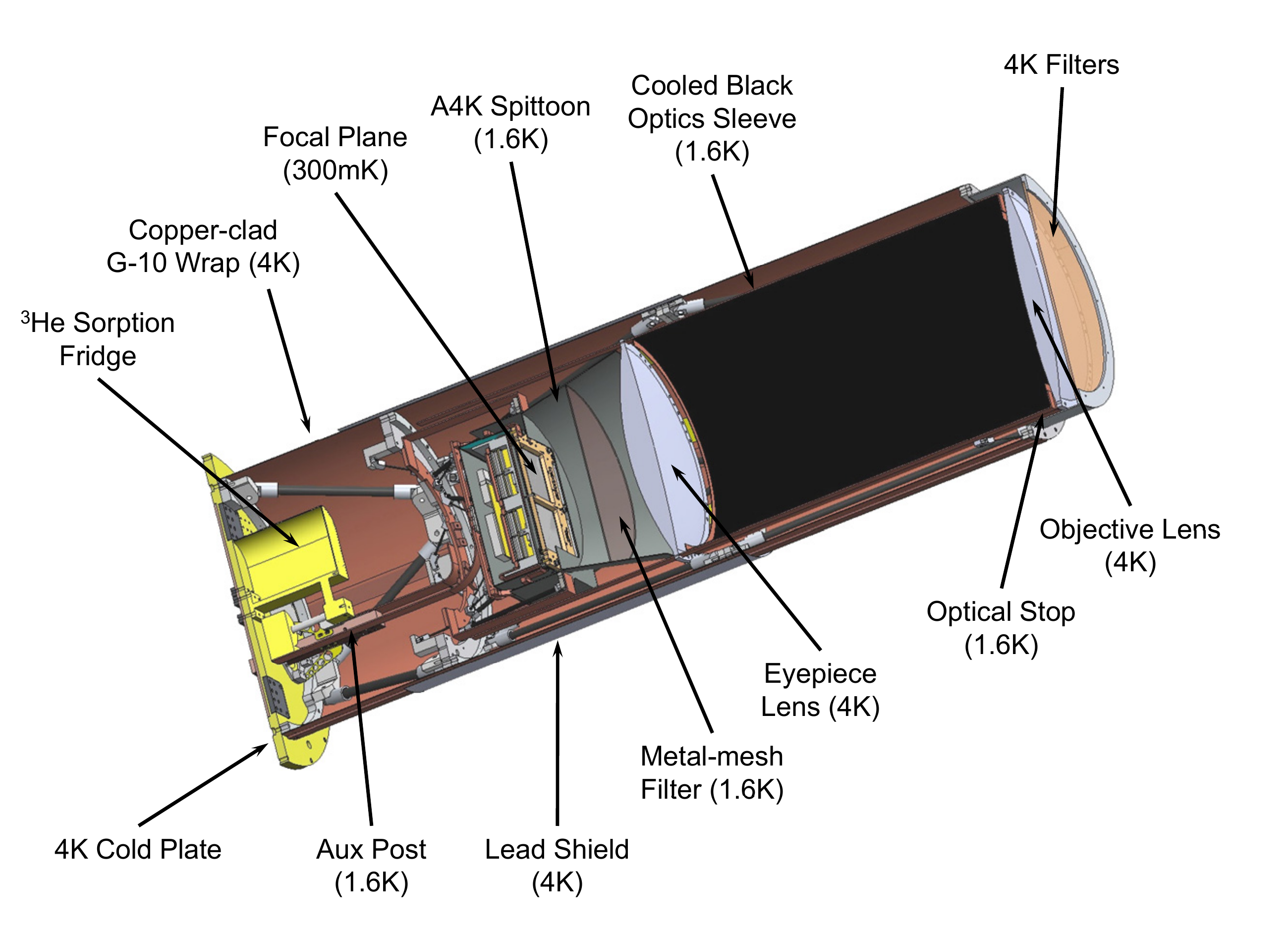}
\includegraphics[height=3in]{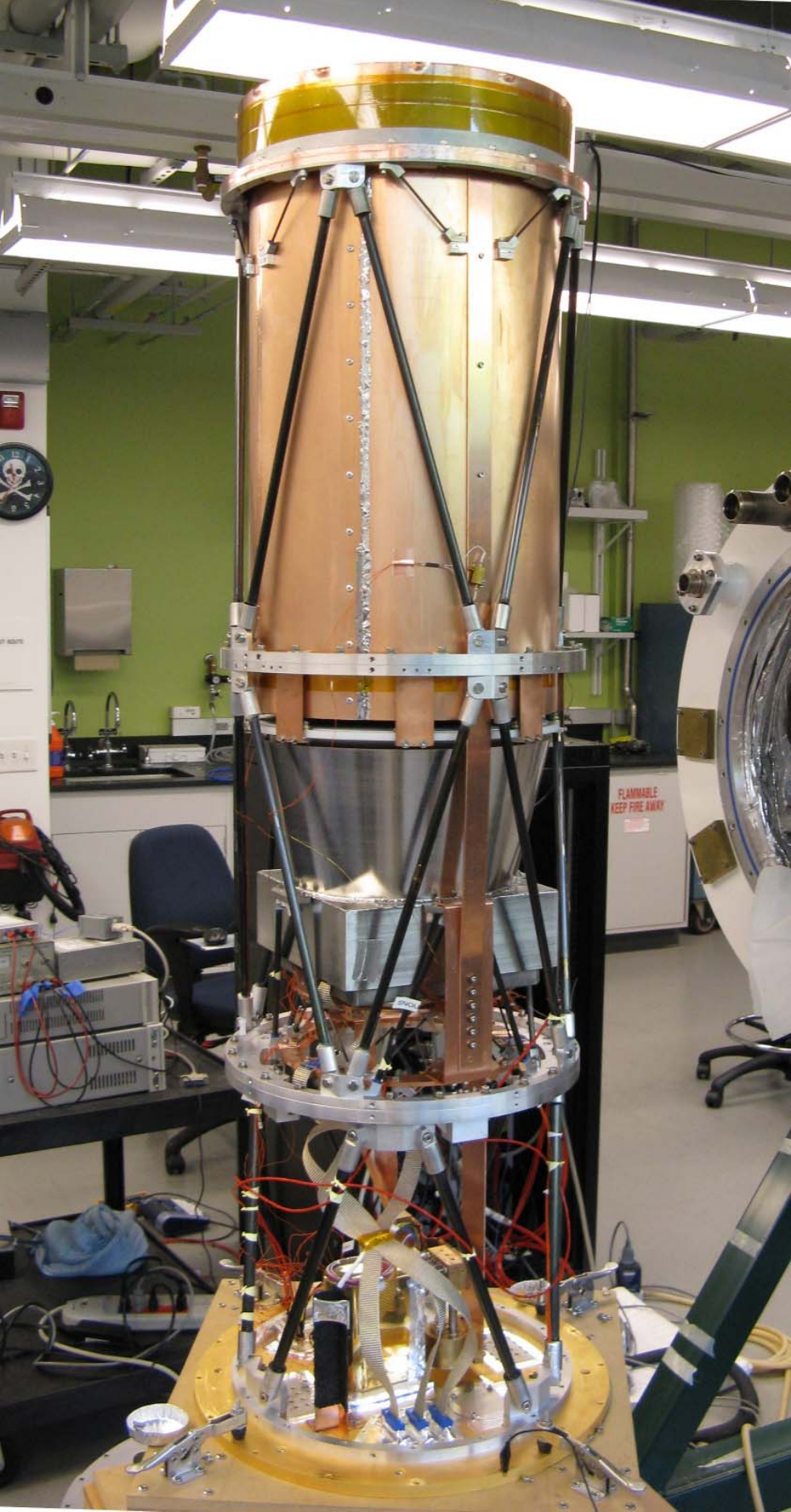}
\end{center}
\caption{\small \emph{At left:} Cross-section of the \spider instrument insert Solidworks model with key components labeled.  \emph{At right:} Photo of the \spider instrument insert without the outermost copper-clad G-10 wrap.  Visible in the photo are the carbon fiber trusses, the high-$\mu$ magnetic shield called the spittoon, the cooled optics sleeve between the lenses, thermal straps, cables, and the cold plate. 
}\label{fig:insert}
\end{figure}

\subsection{Optics Truss Structure and Cold Plate}

A gold-plated $1/2''$ thick $\varnothing17.35''$ 1100-H14 aluminum cold plate forms the base of each \spider insert and is the interface to which the inserts mount to the cryostat helium tank.  This aluminum plate has been light-weighted significantly.  On top of this cold plate is mounted a ${}^3$He sorption fridge and the truss structure that supports the optics and focal plane.  A gold-plated C10100 copper bus bar is mounted to the bottom side of the cold plate and the sorption fridge is bolted to this.  The other side of this thermal bus is connected to copper tabs embedded in the liquid helium tank of the cryostat.  A gold-plated copper bar penetrates the cold plate; the bottom side is connected to the pumped helium auxiliary tank of the cryostat and the other provides a 1.6~K point for the condensation point of the sorption fridge, a thermal intercept ring in the focal plane unit (FPU), a high-$\mu$ magnetic shield, and a cooled blackened light baffle, all of which are described below.

The focal plane unit is supported above the cold plate by a hexapod carbon fiber truss with $\varnothing1/2''$ rods called the camera truss.  The two lenses are then supported from the top of the camera truss by two octopod carbon fiber trusses with $\varnothing3/8''$ rods.  The individual carbon fiber legs have aluminum end caps epoxied to each end with Stycast 2850 and we have used fishing line to center the carbon fiber rods in the end caps to avoid possible galvanic corrosion from electrical contact.  The mounting holes in the end caps have all been reamed to have a slip fit tolerance over $\varnothing1/4''$ shoulder bolts.  The mounting pads on the feet that mount the rod assemblies to the aluminum rings are all coplanar for each leg.  This allows the length of any truss to be changed simply by swapping out legs of different length.  The insert volume in the single-insert test cryostat used for development and characterization is longer than that in the flight cryostat and we will compensate by shrinking the length of each camera truss by $4''$ before installing the inserts into the flight cryostat.

All of the carbon fiber trusses are wrapped in thin copper-clad G-10 sheets to form a light and RF tight sleeve around the inserts.  We have the option of coating the inside of these wraps with a flexible blackening material made from silicone RTV, carbon lampblack, and 316 stainless powder.  We measured the mass loss during cure and heating of many commercially available silicones and found Dow Corning 748 to have very low mass loss.  We have run our cryostat with $\sim1.5$~m${}^2$ of this material and had no issues with out gassing.  Outside of the wraps run four C10100 copper bars to cool the optics trusses.  Each insert is capped with a 1100 aluminum snout that holds a set of filters in front of the objective lens.  These filters are described in section \ref{sec:filters} below.

\subsection{Optics}

\subsubsection{Design Description}
Each \spider insert is a cooled two-lens refracting telescope.  The lens design is identical to that of the BICEP2 experiment and is described in detail in Aikin \etal \cite{aikin:spie2010} (in these proceedings).  Both sides of the two lenses are simple conics and are separated by 550~mm with an effective focal length of 583.5~mm.  This yields a plate scale of 0.98~deg/cm on the focal plane.  The optical system is telecentric to accommodate the flat focal plane geometry.  The field of view of each insert is $20\deg$ across the diagonal.  The entire telescope optics are cooled to 4.2~K to reduce the in-band loading from loss in the thick dielectric lenses.

\subsubsection{Lenses, AR Coats, and Flexures}

The lenses used in \spider are machined from $2''$ thick cast HDPE slabs.  The lens fabrication process involves a number of steps designed to reduce the stress in the lens and achieve the desired lens figure.  Before machining, the HDPE slabs are annealed in a temperature-controlled oven between metal plates and then rough cut to shape.  The rough cut lenses are then clamped in an annealing jig and re-annealed before finish machining on a CNC mill.  The finished lenses are then measured on a CMM to assure the lens figure and surface finish have been archived.  The maximum allowable deviation in the lens figure is $0.005''$ and the lens surface finish is $<63\mu ''$.  The final lenses are $12.4''$ in diameter with an optically-active diameter of $11.4''$.  

Because the index of refraction of HDPE in the millimeter is $\sim1.52$ \cite{lambmaterials}, the instrument would suffer reflections of order $[(n_1-n_2)/(n_1+n_2)]^2 \sim 4\%$ at each lens surface.  To reduce these reflections we bond a tuned layer of porous PTFE sheet to each lens surface manufactured by Porex.  We have chosen a material with a density that yields an index of refraction close to the ideal of $n_{AR} = \sqrt{n_{HDPE}}$ and we order the sheets in thicknesses of $\lambda/4n_{AR}$.  The toleracing of the AR material from the manufacturer is only good to a few mils.  But modeling of the transimission of HDPE (and nylon) slabs with AR coats of various thickness and indicies of refraction shows that $>99\%$ transmission is achieved with thickness and density variations within manufacturing tolerances.

The plastic lenses contract during cooling substantially more than the aluminum mounts from which they are supported ($\sim 2\%$ for HDPE and only $\sim0.4\%$ for aluminum).  The radial differential contraction between the lens and support ring is $\sim0.1''$ at the edge.  To allow for this contraction while keeping the lens well centered and rigid we use eight $1/32''$ thick copper flexures spaced equally around the lenes.  The copper flexures also provide the cooling path for the lenses.  However, the cooldown time of the lenses is limited by the internal thermal time constant of the HDPE.  The length of the flexures are tuned so that the spacing of the lenses is correct after thermal contraction as the insert cools.  Most of the support structure is carbon fiber, so the contraction in the lenses themselves dominates this correction.

\subsubsection{Optical Filtering}
\label{sec:filters}

\spider employs multiple filter types to reduce IR loading on the various thermal stages. We use a stack of four metal-patterned mylar ``IR shaders" on both of the vapor cooled shield (VCS) stages to reduce the IR loading on the helium stage \cite{ade:spie2006,tucker:spie2006}.  We also use hot-pressed resonant metal-mesh filters on the cold side of the 20~K VCS stage and the entrance to the insert, as well as just above the focal plane at 1.6~K.  The entrance to the insert has an AR-coated sheet of $3/32''$ nylon on the cold side to help absorb additional IR.

\subsubsection{Baffling and Optical Stop}

The space between the two lenses is filled with a blackened sleeve that is capped with an annulus that defines the optical stop of the system $9.5''$ in diameter just behind the objective lens.  The stop limits the spillover of the beam onto warmer stages upstream along the optical chain, including the waveplate, filters, and window.  The square phased-array antennas produce a 2D sinc beam pattern and $\sim 25\%$ of the beam power falls outside of the stop.  The blackened sleeve absorbs this sidelobe power and prevents it from escaping out the front of the cryostat.  Because a significant fraction of power is incident upon this blackened sleeve, we susped is from the front of the optics tube using a carbon fiber truss and cool the entire tube to 1.6~K using the pumped helium auxilliary tank (described below).  The sleeve is formed by soldering a tube of copper-clad G-10 and then lining the sleeve with a mixture of Stycast 2850, carbon lampblack, and 316 stainless powder.

\subsubsection{Waveplate}

To increase polarized angle coverage and mitigate the effect of beam
asymmetries, polarization modulation in \spider is achieved via an
AR-coated sapphire half-wave plate (HWP) mounted at 4~K skyward of the
primary optic of each insert~\cite{bryan:spie2010}. The HWP will be
periodically rotated to different angles throughout the flight with a
cryogenic stepper motor. Each HWP will be optimized for the single
frequency band of the 90, 145, or 280~GHz insert in which it
is mounted. A fused-quartz AR coat will be applied to each side. We
have measured the millimeter-wave transmission spectra of birefringent sapphire at
room and liquid helium temperatures in the lab, which are consistent
with our physical optics model. We have also taken preliminary spectra
of an AR-coated 145~GHz HWP integrated with the \spider optics and detector system
in the prototype \spider receiver. Preliminary results show good performance.

\subsection{Focal Plane}

The development of the \spider focal plane was a combined effort with the BICEP2 project.  Orlando \etal \cite{orlando:spie2010} (in these proceedings) describes that effort to develop these platter-style focal planes in detail (these cover revisions A through D).  During the course of testing we found that we could not adequately shield the system from magnetic fields at a level where we would not have to worry about signals induced by spinning the cryostat in earth's field.  This was a fundamental limitation of the focal plane architecture that places the detector tiles and SQUID multiplexer chips on the same plane without the ability to surround the SQUIDs with magnetic shielding.  We have opted to redesign the focal plane and move all of the SQUIDs inside of a closed superconducting niobium box with secondary high-permeability and superconducting shields within.  This focal plane architecture is described in detail below and is referred to as RevX.   
  
\subsubsection{Architecture}

\begin{figure}[!t]
\begin{center}
\includegraphics[height=2.5in]{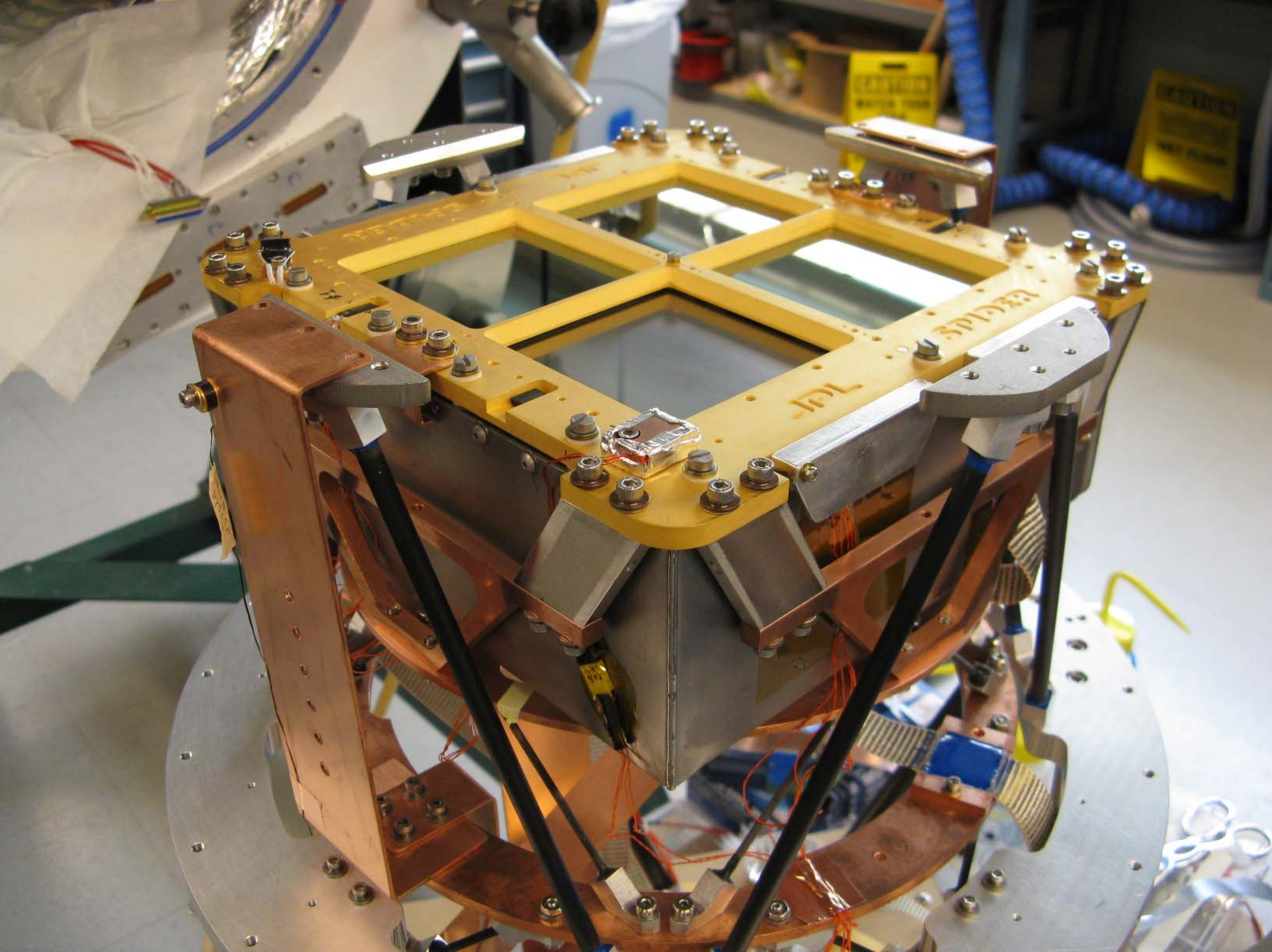}
\includegraphics[height=2.5in]{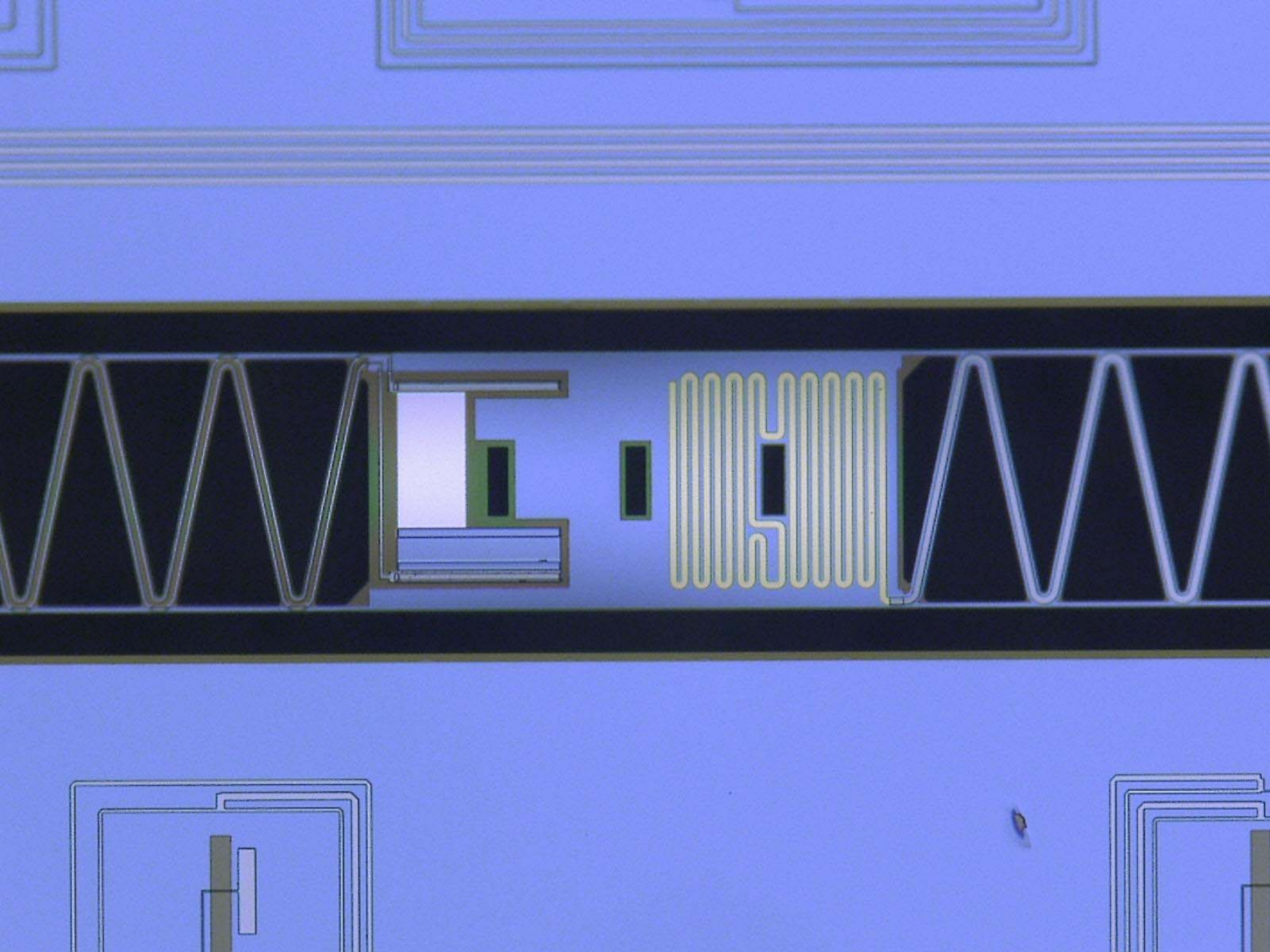}
\end{center}
\caption{\small \emph{At left:} Image of the Spider RevX focal plane without the outer A4K high-$\mu$ shield.  This focal plane has detector tiles in 3 of the 4 positions.  Visible in the photo are the diagonal 316 stainless passive thermal filters, square niobium box shield, and carbon fiber truss and copper thermal straps for the high-$\mu$ shield. \emph{At right:} Close-up image of one of the Spider TESs islands at 400X zoom (photo credit: A. Turner \& J. Brevik).  The meandered legs allow for the requisite low thermal conductivity bolometers to fit in a smaller island cut out.  On the right hand side of the island is the gold resistive meander where the antenna summing tree is terminated.  On the left part of the island are the Ti and Al TESs in series with readout leads leaving the island to the left.}
\label{fig:focalplane}
\end{figure}

The \spider RevX focal plane architecture is shown in figure \ref{fig:focalplane}.  The detector tiles are mounted onto a square gold-plated C10100 copper detector plate $8.55''$ across.  The focal plane structure is cooled to 300~mK with the ${}^3$He sorption fridge.  Pins in the detector plate and slots in the silicon detector tiles register the detector tiles and allow for differential thermal contraction between the silicon and copper.  The detector tiles are held in place with beryllium copper clips.  The detectors are patterned onto the back side of the silicon tiles with respect to the incident light and NSG-N quartz anti-reflection tiles are placed on the front side of the detector tiles to minimize surface reflection from the silicon.  The detectors are spaced $\lambda/4$ from a niobium backshort plate.  The spacing between the detectors and backshort is set using lapped Macor ceramic spacers around the perimeter of the focal plane with a single spacer in the center.  Custom made niobium fasteners attach the copper detector plate to the niobium backshort and the holes in the copper plate are over sized to allow for differential thermal contraction between the copper and niobium.

Behind the niobium backshort is a $1/32''$ thick welded niobium box that encloses all of the SQUIDs.  The signals from the detector tiles are brought around from the detector plate to the inside of this box using flexible aluminum superconducting circuits (flexi circuit).  The aluminum flexi circuit has a measured superconducting transition temperature of 775~mK.  The other side of the flexi circuits are connected to the input stages of the SQUID multiplexing system, which is described below.  Each detector tile has an individual flexi circuit that can carry 128 TES signals.  The flexi circuits are made in two parts which overlap at the center to minimize the size of the slot in the niobium box required for their passage; the effectiveness of the niobium box is limited by the width of these slots.  The niobium box is fastened to the niobium backshort with $40\times$ 7075 aluminum fasteners with a measured superconducting transition temperature of 900~mK.  The total mass of the 300~mK focal plane structure is 8~kg.		

\subsubsection{Detectors and band definition}

The antennas and detectors used in \spider are described in Kuo \etal \cite{kuo} as well as Orlando \etal \cite{orlando:spie2010} (in these proceedings).  A focal plane consists of four detector tiles, each of which has an 8$\times$8 array of dual-polarization pixels (128 detectors) for 145 and 280~GHz, and a 6$\times$6 array (72 detectors) at 90~GHz.  Each pixel consists of two arrays of polarized slot antennas (called A and B and are rotated 90 degrees w.r.t. each other), each of which are summed in phase on superconducting microstrip.  This microstrip passes through a resonant filter that defines the millimeter-wave frequency band and then the signal is fed into a resistive gold meander located on a suspended bolometer island (see figure \ref{fig:focalplane}).  Also on each island are both a titanium and aluminum transition edge sensor (TES) in series.  The millimeter power dissipated in the gold meander heats up the bolometer island and the voltage-biased TES responds to that change in temperature with a change in resistance, thus producing a measurable change in current.

The thermal conductivity ($G$) of the \spider bolometers is low (20~pW/K at $T_C$) to take advantage of the decreased optical loading from a stratospheric balloon platform.  The $G$ of the bolometers are tuned to put the titanium TES ($T_C\sim500$~mK) on transition at float with a margin of safety of $\sim3$.  These low-$G$ devices are saturated on the titanium transition when doing testing in the laboratory, so the aluminum TES ($T_C\sim1.3$~K) in series is used.  This allows characterization of beams and frequency bands in the lab before flight.  The TES bolometer islands have some response to high-frequency ``blue leaks" since the microstrip running to the island acts like an antenna within the cutout of the niobium ground plane.  In an effort to push that frequency response above our filter cutoff we have shrunk the window cutout of the islands.  The lower $G$s of the \spider devices required that the island legs that determine the device $G$ be meandered to obtain enough length for the required thermal conductivity.  

\subsubsection{SQUID multiplexer and cabling}

At the heart of \spider's signal readout chain is a three-stage SQUID time-domain multiplexer made by NIST \cite{deKorte:2003mux}.  SQUIDs are remarkably sensitive to changes in magnetic flux, so coupling an input coil into a SQUID loop forms the basis of a very sensitive current readout. Each bolometer signal is fed through a low-pass $L/R$ filter that reduces aliased noise and then into the input coil of a clover leaf SQUID (SQ1).  A single mux chip handles the signals from 32 TESs and a ``dark" SQUID (used to monitor noise and magnetic field pickup) and combines these signals into a summing coil that is fed into a second stage SQUID (SQ2).  The individual SQ1 are biased in sequence, so the summing coil (and hence, SQ2) only has the signal from one of the TESs at any time.  The SQ1 bias lines are referred to as ``row selects" because they select which SQ1 is active.  The SQ2 signal is then fed into a SQUID series array amplifier (SSA) that has 100 SQUIDs in series per channel.  The timing, biasing, and readout of the SQUIDs is controlled by the Multi-Channel Electronics (MCE) crate made by UBC \cite{mce}.  Each MCE crate handles the 512 detector signals from an entire insert with only a power cable and fiber optic pair leaving the crate.

As mentioned above, the signals from the detectors are routed to the SQUID chips on superconducting flexible circuits.  The multiplexer chips are mounted on circuit boards with copper traces onto which superconducting PbSn solder has been electroplated between the mux chips and the bond edge.  Each of these ``MUX boards" has four sets of SQUID chips which are enough to handle the signals from one detector tile (up to 128 devices).  The typical impedance of the TESs on the titanium transition is $\sim30-40$~m$\Omega$ and the TES bias shunt resistors are only 3~m$\Omega$.  The superconducting aluminum and PbSn circuit between the TESs and mux chips reduce the parasitic impedance along this path to a level well below the shunt resistance. 

The signals from the SQ2s, SQ1 row select lines, detector biases, and SQUID feedback lines are feed to Samtec headers at the edge of the MUX boards and the four MUX boards plug into one breakout board which routes the signals to four 100-way micro-D connectors.  All of the outputs of the SQ2s are run through one of these 100-way connectors up to a separate circuit board that houses the two SSA modules.  The other three 100-way connectors have all of the signals, biases, and feedback lines and run to the cold plate on NbTi ribbon cables.  There are 300 wires leaving the focal plane and the detector bias and SQ2 bias lines can carry currents up to a couple mA each.  The joule power dissipation by manganin cables would have been too much for our ${}^3$He fridge to handle.  So we have chosen to make all of the wires running to the focal plane superconducting (with the exception of some thermometry).  The cables running from the 4~K cold plate to the 300~K vacuum feedthrough to the MCE crate are all teflon-jacketed shielded twisted-pair manganin.

\subsubsection{Magnetic shielding}

A serious concern to the \spider project is magnetic field pickup in the system that cannot be separated from astrophysical signals.  One of the strengths of the \spider instrument is the ability to make long scans across the sky and measure signals on large angular scales.  The earth's magnetic field forms a dipole that we will scan through and magnetic pickup in the various SQUID stages and the TESs themselves will be difficult to discriminate from the desired signal.  Changing magnetic flux through the SQUIDs produces a shift in the $V-\phi$ curve.  To reduce this effect, NIST has fabricated the SQUIDs in a counter-wound cloverleaf pattern so that they are nominally only sensitive to 2nd order gradients in magnetic field.  Changing magnetic flux will also induce a current in the superconducting summing loop that combines the signals from all of the SQ1s and feeds it into SQ2.  This loop area has been minimized but is still finite.  The exact value of $T_C$ of the TESs depends upon the magnetic field and so a change in magnetic environment will shift $R(T)$ and produce a change in current.  

To that end we have tried to shield the focal plane from magnetic fields so that pickup from the earth's field will be below the expect signal from the CMB.  Our shielding scheme consists of multiple layers of both high-permeability and superconducting materials.  The TESs are spaced $\lambda/4$ away from the large superconducting niobium backshort and the entire focal plane is surrounded by a large high-permeability ``spittoon" made of Amuneal A4K material to reduce changes in $T_C$ of the TESs.  As mentioned above, the TES signals are routed inside of a superconducting niobium box that is closed with superconducting aluminum fasteners.  We have tried to minimize the size of all slots penetrating the niobium box by overlapping the flexi circuits to halve their width as well as truncate the two thermal straps where they enter the box.  Immediately inside the niobium box at the level of the slots is another high-permeability A4K plate that wicks entering magnetic field away before it can get to the SQUIDs.  There is a reentrant superconducting 1100 aluminum open-top box that the flexi circuits run up and over where the SQUID mux chips are located.  All of the SQ1 and SQ2 chips (which include the superconducting summing coils) are further shielded with squat, open-ended sleeves that surround the mux boards (each has four sets of mux chips).  The SSA boxes are suspended in the aluminum box and are housed in compact open-ended niobium boxes with Cryoperm sheets inside.  These niobium boxes are further wrapped in ten layers of 0.6~mil Metglas 2714A.  Outside of the insert we have wrapped a $20''$ long, $0.006''$ thick superconducting lead sleeve that is centered on the focal plane structure.  Lastly, each of the insert tubes is lined with a two-layer $0.040''$ thick A4K shield that runs the length of the helium tank.

\begin{figure}[!t]
\begin{center}
\includegraphics[height=2.25in]{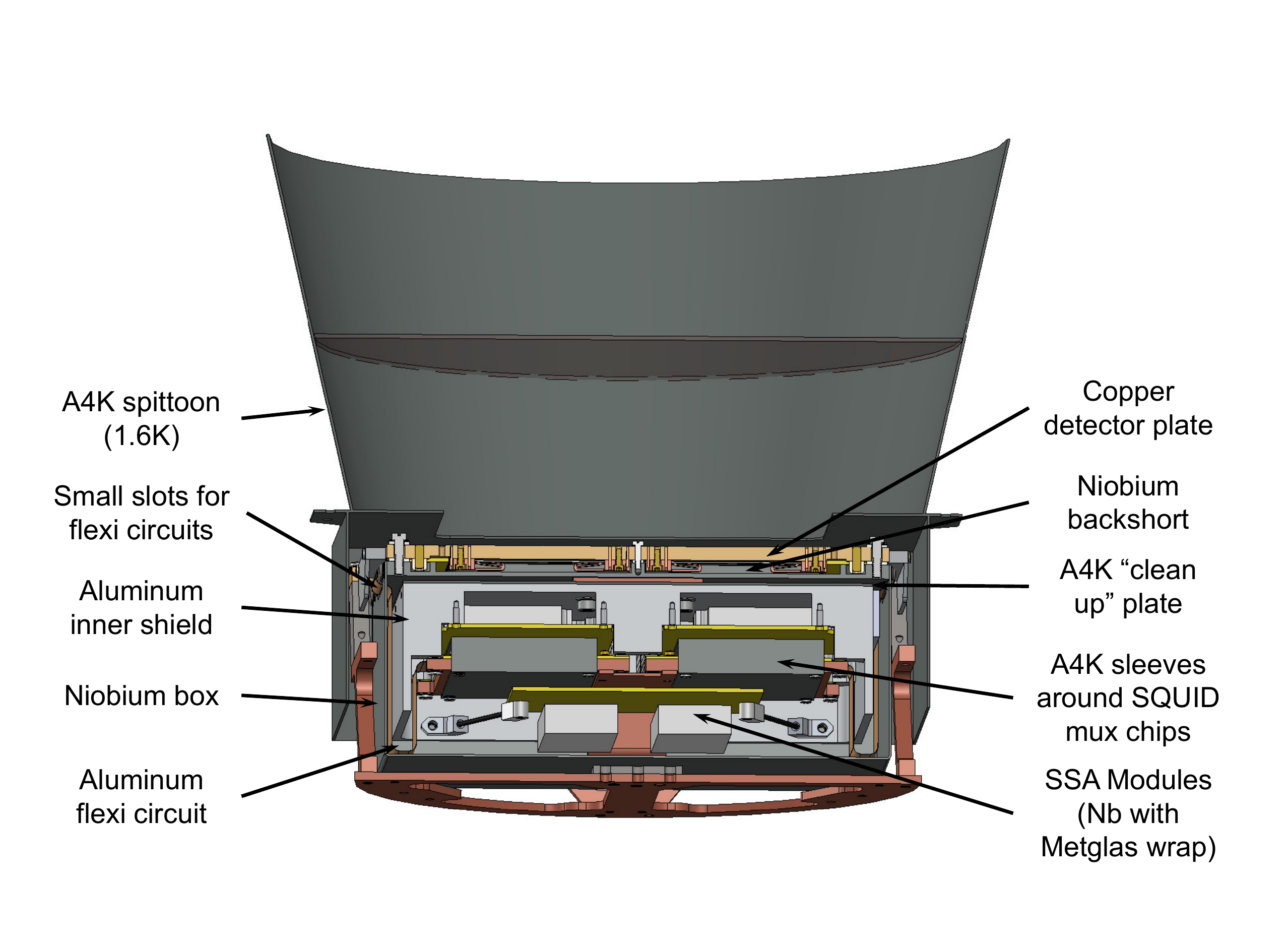}
\includegraphics[height=2.25in]{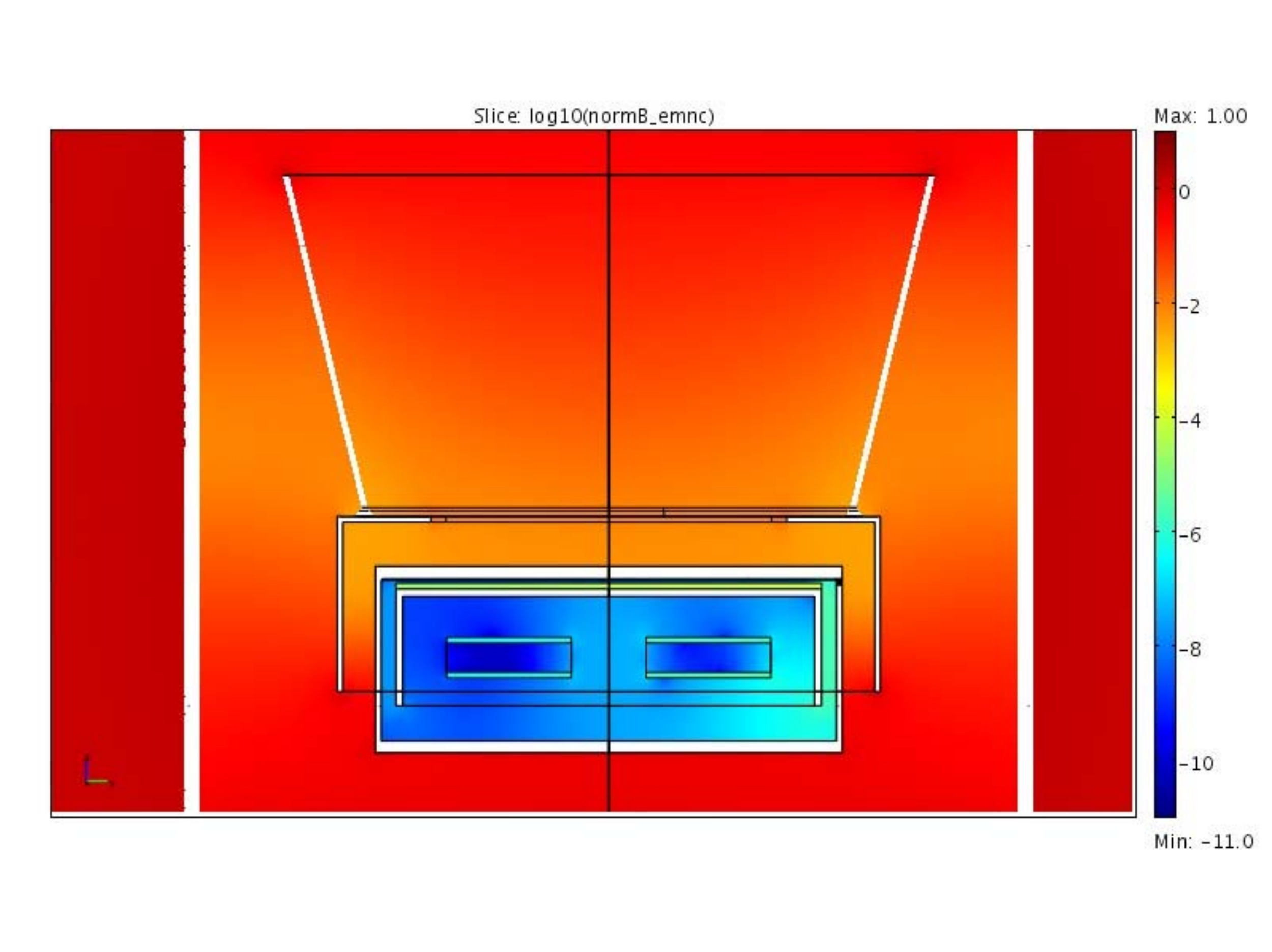}
\end{center}
\caption{\small \emph{At left:} Cross-section schematic of some of the key magnetic shielding components of the \spider RevX focal plane assembly.  Everything shown except for the ``spittoon" sits at 300~mK.  \emph{At right:} Results of a magnetic field model of the focal plane structure using COMSOL to determine the shielding factor at the SQUID multiplexer chips (located within the inner-most rectangles).  The model and results are described in the text.}
\label{fig:magshield}
\end{figure}

We have modeled the magnetic shielding configuration using COMSOL Multiphysics.  This model does not have the outer-most Cryoperm-10 shield to keep the mesh size manageable.  The results of the model for a 1~T incident magnetic field are shown in figure \ref{fig:magshield} where we display the $log_{10}$ of the resulting field amplitude.  COMSOL does not handle multiply-connected superconducting regions properly.  So we model the effects of the narrow $0.10''\times1.35''$ slots in the niobium box through which the flexi circuits pass one at a time. The direction of the magnetic field is into the page which is along the long axis of the slots (the direction of maximum magnetic field penetration).  This model is idealized but suggests that the shielding configuration of the RevX focal plane is capable of yielding shielding factors in excess of $10^8$ at the SQUIDs, not including the additional A4K tubes that run the length of the helium tank and are expected to provide an additional shielding factor of $\sim50$.  We have estimated the shielding factor to earth's magnetic field necessary for the flight to be $\sim10^7$.

\subsection{Cryogenics}
\label{sec:cryogenics}

\subsubsection{Overview}

The \spider Long Duration Balloon (LDB) cryostat and cryogenic systems are described in detail in Gudmundsson \etal \cite{gudmundsson:spie2010} (in these proceedings).  The cryostat holds $\sim1000\ell$ of liquid helium in a tank shaped like a 6-shot revolver cylinder.  There are seven holes in the tank; the six around the outside house the six instrument inserts and the central hole is used to pass wiring up to the waveplates and thermometry in front of the inserts.  The six instrument cold plates (described above) bolt to the liquid helium tank via gold-plated aluminum interface plates.  In addition to the main helium tank, a smaller $20\ell$ auxiliary tank (Aux tank) of helium is fed via a capillary tube from the main tank and is pumped to 1.6~K.  On the ground, this pumping is achieved with a vacuum pump but in flight the reduced atmospheric pressure at float provides the pumping.  As described below, the Aux tank provides an intermediate temperature stage between our sub-K fridge and the main liquid helium bath. 

\subsubsection{Truss structure and thermal management}

The focal plane unit (FPU) is supported on eight 316 stainless ``heat capacity blocks".  The cooling path from the FPU to the ${}^3$He fridge still runs through these blocks and we have measured them to have a combined thermal impedance of 2.3~mK/$\mu$W, in excellent agreement with COMSOL simulations.  They are gold plated on the ends to minimize thermal boundary impedance.  Stainless steel has a relatively high heat capacity and these blocks act as a passive thermal filter to temperature fluctuations generated below in the truss structure, thermal straps, or the fridge.  We have measured the thermal transfer function of the passive thermal filter to have a 3dB point of 2~mHz. 

The heat capacity blocks connect the four sides of the FPU to a copper ring via copper supports.  This whole structure is cooled to 300~mK and is referred to as the sub-Kelvin stage (sub-K stage).  The sub-K stage is supported off of another copper ring cooled to 1.6~K with the pumped helium auxiliary tank with carbon fiber rods.  This ``aux ring" is itself supported from a 4.2~K aluminum ring with carbon fiber rods.  We measured the thermal conductivity of many polymeric and composite materials during the course of the development of the \spider truss structure and found carbon fiber rods to have the highest ratio of elastic modulus to thermal conductivity of non-brittle materials over the range of temperatures of interest \cite{runyan08}.  The truss structure appears gossamer but is very stiff.  We have modeled the deflection with COMSOL and expect the 8~kg focal plane to deflect a few mils over a $90\deg$ tilt, corresponding to much less than an arcminute of beam deflection on the sky.

There is a concern that the vibration of rigid metal thermal straps attached to the cold stages might vibrate at frequencies within the science band and couple power into the detectors.  The stainless heat capacity blocks described above are part of the strategy to mitigate that effect. We have opted instead to use flexible thermal straps formed with many layers of $0.001''$ C11000 copper foil that have been electron beam welded into copper mounting tabs.  Despite the foil alloy being C11000, we have measured its RRR to be $\sim200$ and confirmed that they have correspondingly high thermal conductivity.



We use a combination of DT-series silicon diodes and Cernox resistance thermometers (both made by Lakeshore Cryotronics) to monitor the temperatures within the cryostat and inserts.  We exclusively use resistance thermometers at all sub-K stages and diodes for all stages with temperatures greater than 1~K.  

\section{\spider System Performance}

\subsection{Optical Performance}

\spider will observe in three frequency bands at 90, 145, and 280~GHz.  Most of the insert development effort has focused on the 145~GHz system, and consequently, it is the most well characterized of our frequencies.  Here we present some of the optical performance results from the 145~GHz system.

\subsubsection{Bands}

  As mentioned above, \spider's frequency bands are defined with resonant on-tile filters between the detector antenna and the resistive meander on the TES island \cite{kuo}.  We have measured the spectal response of one of the inserts configured for 145~GHz operation.  The average bandwidth for the 145~GHz channels is 34~GHz (using the convention of Runyan \etal \cite{runyan03}), or 25\%.  Even at float we are concerned about residual atmospheric emission and have worked to fit the frequency bands within the available atmospheric windows, paying particular attention to avoiding water lines.  

\begin{figure}[!t]
\begin{center}
\includegraphics[width=6.5in]{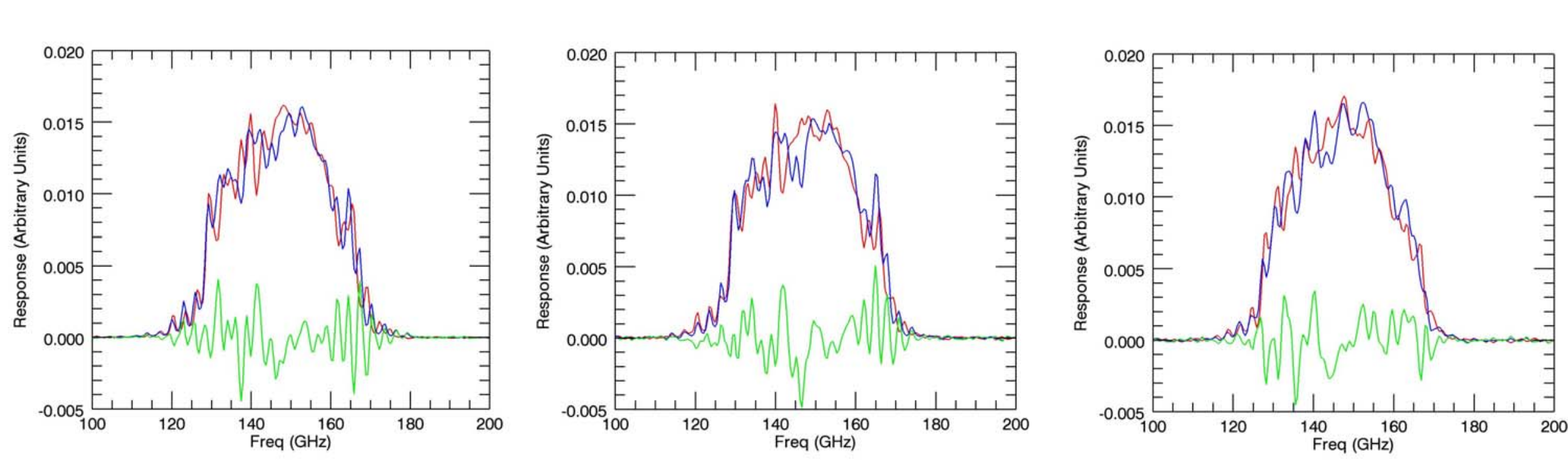}
\end{center}
\caption{\small Measured signal bands from three 145~GHz pixel pairs with high S/N.  The red and blue curves show the bands for each of the two antennas in a pixel pair and the green curve is the difference between the two.  The bands have been normalized to have the same integrated value.}
\label{fig:bands}
\end{figure}

\spider's observing strategy will involve differencing the two polarizations in a given pixel and a mismatch in spectra could limit the effectiveness of this strategy.  Figure \ref{fig:bands} shows examples of measured spectral bands at 145~GHz.  In the figure, the spectral bands have been normalized to have the same integrated value over the band.  The oscillatory nature of the difference band suggests that common mode emission sources with smoothly-varying spectra will be well-differenced.  The \spider filter strategy also effectively removes high-frequency spectral leaks that could couple high-frequency power onto the detectors.  We have measured the high-frequency response of our 145~GHz band using a chopped thermal source and a thick-grill filter with a cut off of 185~GHz and found the response to be less than a few tenths of a percent of the in-band signal (consistent with the noise floor of the measurement).

\subsubsection{Optical efficiency}

We needed a way to characterize the \spider instrument inserts under flight-like loading conditions.  To do this we developed a helium cold load cryostat that bolts to the front of either the insert test cryostat or the flight cryostat and presents a cold, millimeter-black, beam-filling surface to the instrument.  This cold load cryostat is a $23\ell$ helium cryostat with a $15.5''$ cold plate.  Using a carbon fiber truss, we stand off a copper plate covered in millimeter-black pyramidal tiles manufactured by Thomas Keating, Ltd.  Embedded within these tiles are multiple Lakeshore SD diodes to monitor the load temperature.  We have mounted a resistive heater on the back side of the black load and can elevate it to any temperature above 5~K.  The cold load cryostat has radiation shields attached to the helium bath and its vapor cooled shield and we have mounted IR-blocking filters to each of these stages to reduce the optical loading on the black absorber.

A convenient feature of being able to elevate the cold load temperature is that we can measure the end-to-end optical efficiency of our system by measuring the optical loading on the detectors, $P_{opt}$, as a function of cold load temperature, $T_{CL}$.  The only optical element missing in this configuration is the cryostat vacuum window (the cold load and main cryostats share a common vacuum) and we have added a handful of IR filters on the cold load side which will not be there in flight.  We ramp the cold load temperature from 5~K to $\sim20$~K and take load curves to measure the loading at each temperature.
 We convert the actual cold load temperature into an equivalent Rayleigh-Jeans temperature, $T_{RJ}$, and then take the slope $dP_{opt}/dT_{RJ} = k\int\eta_\nu~d\nu$, where $k$ is Boltzmann's constant, $\eta_\nu$ is the measured spectral response normalized by the optical efficiency, and we have assumed single-moded performance ($A\Omega=\lambda^2$).  We measure $dP_{opt}/dT_{RJ}$ to be $\sim0.17$~pW/K$_{RJ}$ at 145~GHz.  We can then calculate a band-average optical efficiency $\overline{\eta} = (\int{\eta_{\nu}~d\nu})/\Delta\nu$.  The 145~GHz \spider inserts have a typical end-to-end optical efficiency of 36\%. 

\subsubsection{Internal loading}

To fully take advantage of the stratospheric balloon platform we need to reduce the internal loading to a level where the additional photon noise contribution does not significantly increase the overall system noise.  We have estimated the loading from the CMB and atmosphere at float to be $\sim0.28$~pW per polarization at 145~GHz.  This corresponds to a RJ temperature of $\sim1.7$~K for a system with 36\% optical efficiency and a 34~GHz bandwidth.  The loading from a 300~K vacuum window with 0.5\% loss is $\sim0.2$~pW.  So we would like the loading from within the cryostat to be a small fraction of a pW, or only a few K${}_{RJ}$.  

The cold load provides a convenient way to measure the internal loading from within the cryostat.  We extrapolate the $P_{opt}$ versus $T_{RJ}$ curve described in the previous section back to zero and read off the internal loading as the y-intercept.  Note that this technique does not include the loading from the warm vacuum window, which is estimated to be $\sim0.25$~pW.  Before installing the cooled black optics sleeve between the lenses we measured the internal loading from within the cryostat to be $\sim0.6$~pW, or $\sim3.5$~K${}_{RJ}$ at 145~GHz.  We have since installed the cooled optics sleeve and by measuring the coupling of the detectors to the sleeve we expect the internal loading to decrease by 0.3 to 0.4~pW.

\subsubsection{Beams}

We have also done some preliminary beam characterization at 145~GHz by looking into the laboratory through a $6''$ thick Zotefoam PPA-30 window (this will not be the flight window configuration).  We have made near-field beam maps a few inches away from the window aperture using a chopped thermal source on an X-Y translation stage as well as far-field slices using a large chopped thermal source on a linear stage that can be rotated about the bore sight.  The near field beam maps are in qualitative agreement with physical optics modeling using Zemax and the far-field beam slices for the few pixels measured show well-matched gaussian beam pairs for the two polarization antennae within a pixel with a FWHM of $31'$ at 145~GHz. The difference in the A and B polarizations within an individual pixel are measured to be $<2\%$ of the average beam.  See Aikin \etal \cite{aikin:spie2010} in these proceedings for more details on beams from the BICEP2 optical system, which is very similar to \spider's. 

\subsection{Detector performance}

We have measured detector parameters for a handful of prototype detector tiles.  The \spider-specific engineering tiles have two different values of $G$ so that we can determine the optimum value for the flight. 

\subsubsection{Noise}
			
\begin{figure}[!t]
\begin{center}
\includegraphics[height=2in]{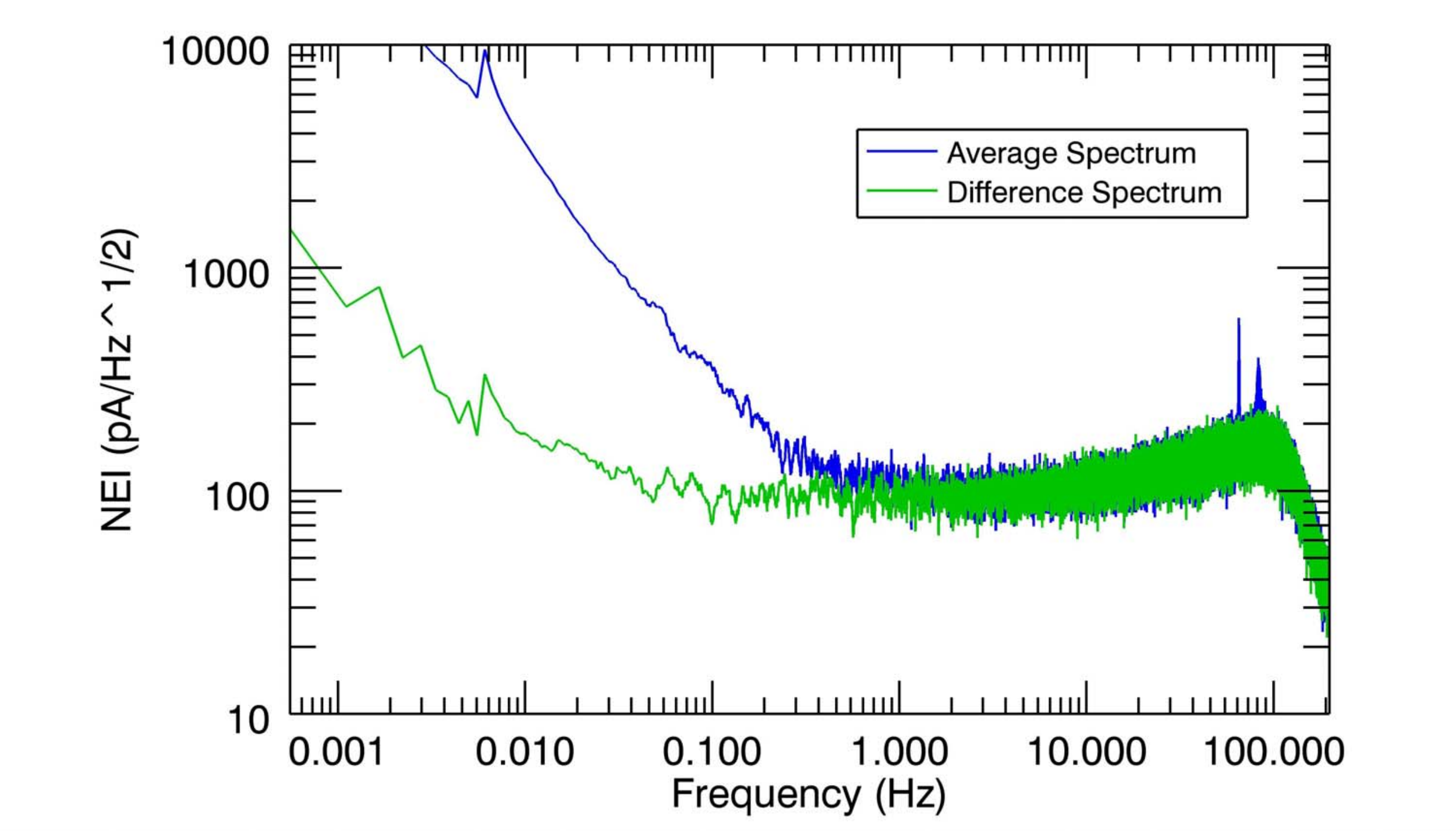}
\end{center}
\caption{\small Measured noise spectra for a pixel pair in noise equivalent current for half-hour time traces with the TESs midway up the transition while staring into a 5.5~K blackbody.  The blue curve shows the noise spectrum of the average, $(A+B)/2$, of the two noise traces from the pixel pair and the green curve is the difference, $(A-B)/2$, of the pixel pair.  Both of these spectra have been smoothed over 20 points. The $1/f$ noise and the common spikes at higher frequency are effectively removed by the difference. }
\label{fig:noise}
\end{figure}

We have measured the instrument noise at 145~GHz versus $R/R_n$ on the TES transition while staring into the cold load cryostat at 5.5~K to simulate flight loading.  We calibrate the noise using bolometer load curves and the measured optical efficiency of the system, $dP/dT_{RJ}$, and then convert the noise PSDs into $NET_{CMB}$.  We also confirm this calibration by varying the temperature of the blackened cold load plate and measure the response on the detectors.  Some example noise spectra from an A/B pixel pair (as well as the difference PSD) is shown in figure \ref{fig:noise}.

The angular scales on which \spider will be able to measure the CMB power spectrum will be affected by the combination of scan speed and $1/f$-noise.  We have measured the $1/f$-knee frequency of undifferenced bolometers to be typically 0.1 to 0.4~Hz.  Taking the difference of the two detectors in an A/B pixel pair removes the common optical and electrical components and lowers the $1/f$ frequency to $\sim60$~mHz.  This corresponds to $\ell$ of a few for the likely scan speeds of \spider.  As seen in the figure, pixel pair differencing produces three decades of clean signal bandwidth. 

\subsection{Magnetic shielding performance}

As mentioned above, the response of the system to magnetic fields is a significant concern to \spider.  To measure the magnetic field response of the system we use two Helmholtz coils 79~cm in diameter.  These coils can be spaced in true Helmholtz configuration for the direction along the optic axis of the test cryostat, but must be spaced further than $D/2$ for the two orthogonal axes due to the diameter of the cryostat.  In all configurations we center the coils about the focal plane and calculate the nominal field strength at the center in the absence of magnetic materials.  The coils are driven with a AE Techron current amplifier with a sine wave frequency generator and can produce a field strength of many 10s times earth's field. 

The principle motivation for converting from the flat focal plane architecture of RevA-C to the shielded box architecture of RevX was to reduce the magnetic field pickup in the system.  We have reduced the magnetic field response by at least an order of magnitude for most channels by converting to the RevX design.  In fact, for the majority of channels we can only place upper limits on the magnetic response.  

We have also been unable to detect significant magnetic response in the TESs themselves due to shifting $T_c$.  Although we haven't seen response of the detectors themselves due to changing magnetic fields, the response of the SQUID multiplexing system will appear as a signal in the detector time streams.  We can calibrate the response of the system to magnetic fields into K${}_{CMB}$ per earth's field ($B_e$) and find that we can place a limit of the magnetic field response of $<10~\mu K_{CMB}/B_e$ in the majority of channels at 145~GHz for all three axes without any attempt to remove pickup (\eg with A/B pixel differencing or deprojection of the non-antenna coupled SQUID in each MUX column).  In previous focal plane shielding iterations we found that pixel pair differencing reduces the magnetic signal by 1 to 2 orders of magnitude.  Even at this limit, the signal from earth's field would be much less than the CMB dipole but would still be detectable in \spider's CMB maps.  We hope to do longer integrations at high magnetic fields to put even tighter limits on the magnetic response.


\acknowledgments      

The \spider collaboration gratefully acknowledges the support of NASA (grant number NNX07AL64G), the Gordon and Betty Moore Foundation, and NSERC.  With great sadness, the \spider collaboration acknowledges the countless contributions of Andrew E. Lange, the late PI of the \spider project.  His wisdom and selfless leadership will be sorely missed.  WCJ acknowledges the support of the Alfred P. Sloan Foundation.  The authors gratefully acknowledge our collaboration with the BICEP2 and Keck projects.  The author thanks J. Lazear for his help with the design and construction of the cold load.

\bibliography{mcrspie}   
\bibliographystyle{spiebib}   

\end{document}